\documentclass[a4paper,twocolumn,11pt]{quantumarticle}
\pdfoutput=1
\usepackage[utf8]{inputenc}
\usepackage[english]{babel}
\usepackage[T1]{fontenc}
\usepackage{amsmath}
\usepackage{hyperref}
\usepackage{cite}
\usepackage{tikz}
\usepackage{lipsum}

\RequirePackage{mathtools}
\usepackage{tikz-cd}
\usetikzlibrary{cd}
\usepackage{mathrsfs}
\usepackage{amsfonts}
\usepackage{eucal}
\usepackage{latexsym}
\usepackage{amsthm}
\usepackage{amsmath}
\usepackage{faktor}
\usepackage{xfrac}
\usepackage{graphicx}
\usepackage{xcolor}
\usepackage[caption=false]{subfig}
\usepackage{dsfont}

\usepackage{skak}
\usepackage{physics}
\hypersetup{colorlinks,%
citecolor=blue,%
filecolor=blue,%
linkcolor=blue,%
urlcolor=blue,%
pdftex}
\usepackage{CJK}

\newcommand{\phantomsubfloat}[1]{
    {
        \captionsetup[subfigure]{labelformat=empty}
        \subfloat[][]{#1}
    }%
}

\begin{document}

\title{A new realization of the long-range entanglement: fractality replacing the topological order}

\author{Wei Wang}
\affiliation{Tsung-Dao Lee Institute, Shanghai Jiao Tong University, Shanghai, 201210, China}
\email{wwwei\_wwwang@sjtu.edu.cn}
\maketitle

\begin{abstract}
The essence of the famed long-range entanglement as revealed in topologically ordered state is the paradoxical coexistence of short-range correlation and nonlocal information that cannot be removed through constant-depth local quantum circuits. Its realization in different quantum states is a focus research topic in both quantum computation and quantum matter. However, the proved realizations are subject to the paradigm of topological order (including its extensions), i.e. via a quantum code structure with macroscopic code distance. Here, we broaden the knowledge of long-range entangled states by rigorously proving the coexistence in a new concrete state. The state describes qudits on the newly experimentally discovered fractal lattice geometry (1.58D) on which the quantum code structure has been shown not to exist, i.e., there is no topological order. Our result might reveal a new paradigm for the realization of the long-range entanglement in many-body quantum states, and might stimulate new studies connecting quantum information and quantum matter.
\end{abstract}

\section{Introduction}
The long-range entanglement (LRE), as revealed in topologically ordered (TO) states, is the quantum-information foundation for the emergent properties, e.g., anyonic excitations, that are vital in quantum computation and distinctive in quantum phase of matter~\cite{Wen1989,Wen1990,WenNiu1990,Moessner2001,Kitaev2003,Freedman2003,Kitaev2006,Chen2010,Isakov2011,Mazac2012,Kim2013,Coldman2016,Wang2017,Shirley2018,Zeng2019,Satzinger2021,Semeghini2021,Erhard2021,Rahmani2020}. It is believed that the peculiarity of LRE is rooted in its paradoxical nature: In a LRE state, certain nonlocal information is encoded so that it cannot be prepared from any product state through constant-depth local quantum circuits; however, the nonlocal information does not manifest in long-range correlations, but dissolves the correlation between nonadjacent local degrees of freedom, resulting in short-range correlation~\cite{Kitaev2003,Chen2010}. The short-range correlation, or even zero correlation length, distinguishes LRE from other nonlocal entanglement structures, e.g., the GHZ-type entanglement, entanglement for critical-point and gapless phases, etc~\cite{Zeng2019}.

LRE attracts growing attention on its realization in different quantum states of specific interests~\cite{Shirley2018,Zeng2019,Manna2020,Satzinger2021,Semeghini2021,Rahmani2020,Zhu2021}. In reality, the paradoxical essence of LRE is only realized in the class of quantum states where the majority possesses manifold-topology characterizations, and 1D states have been completely excluded~\cite{Zeng2019}. Furthermore, the proof for LRE relies on the structure of TO (including extensions of 2D TO in 3D or higher), or equivalently a quantum code with macroscopic code distance, e.g., locally indistinguishable ground states~\cite{Kitaev2003,Levin2005,Bravyi2006,Chen2010,Zeng2019,Guth2014,Shirley2018,Evra2020,Zhu2021}.

The attention on LRE leads to a fundamental question: Is there a new paradigm for realizing the paradoxical nature of LRE? Here, we try to contribute to answering this question. Our approach is to prove LRE in a concrete quantum state where the code structure does not apply. Due to the absence of the characteristics of TO, we expect that the proof signifies the existence of a novel paradigm.

\begin{figure}[ht]
\centering
    \includegraphics[width=8.3cm]{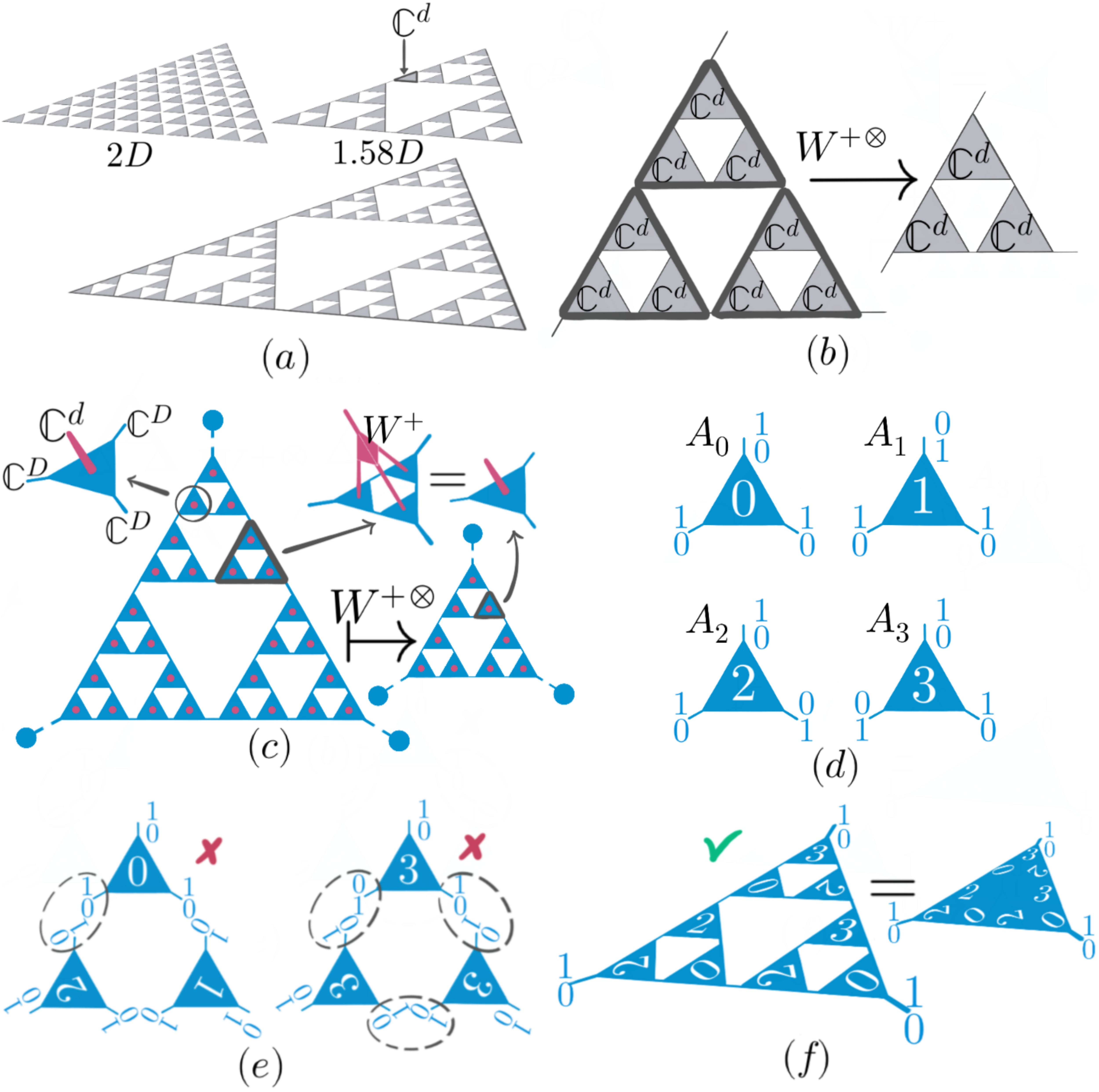}   
\phantomsubfloat{\label{1a}}\phantomsubfloat{\label{1b}}\phantomsubfloat{\label{1c}}\phantomsubfloat{\label{1d}}
\phantomsubfloat{\label{1e}}\phantomsubfloat{\label{1f}}
\vspace{-2\baselineskip}
\caption{(a) The self-similarity (zoom-in and zoom-out invariance) of the Sierpi\'nski-gasket (right) prevents the lattice from reaching continuum limit in 2D (left). (b) The coarse graining based on blocking qudits resembles the zoom-out operation. The dark circulation denotes a block. (c) The tensor network formed by copies of the tensor $A$ and constant corner tensors. The invariance condition of $A$ and the self-similarity of the tensor-network under the coarse graining. (d) Tensors $A_0$, $A_1$, $A_2$ and $A_3$ that sum to $A$. (e) Zero contractions and (f) nonzero contraction of copies from $A_0$, $A_1$, $A_2$ and $A_3$. The circulated are flips between ${1\atop{0}}$ and ${0\atop{1}}$.}
\end{figure}

The concrete state $\ket{\Psi}$ describes qudits ($\mathbb{C}^d$) on the Sierpi\'nski gasket fractal with fractional dimension 1.58D. Recent experimental discovery of quantum matter on this lattice (see Fig.~\ref{1a})~\cite{Shang2015,Wang2018,Kempkes2019,Liu2021,Xu2021,Biesenthal2022} has inspired a range of studies on the existence of LRE and topological properties in fractal quantum states~\cite{Brzezinska2018,Agarwala2018,Pai2019,Fremling2020,Manna2020,Iliasove2020,Yang2020,YangZ2020,MannaDuncan2021,Fischer2021,Sarangi2021,Zou2021,Li2022}. However, the latest studies~\cite{Manna2020,MannaDuncan2021,Li2022,Zhu2021} indicate that topological signatures in fractal states with dimension between 1D and 2D cannot realize LRE. An intuitive reason is the incompatibility between the fractality and the continuum-limit characterization of states (see Fig.~\ref{1a}) which invalidates the manifold-topology picture. More formally, Ref.~\cite{Zhu2021} shows that the incompatibility invalidates the string-like logical operators that are inherent in all 2D TO. Accordingly, Ref.~\cite{Zhu2021} proved the nonexistence of the code structure for defining $\mathbb{Z}_N$ TO and claimed straightforward generalization of the proof to all 2D TO. In other words, in fractal systems that can be embedded in 2D, TO is believed not to exist. Ref.~\cite{Zhu2021} then accordingly speculates the nonexistence of LRE.

It might be unexpected that in the absence of TO, i.e. the code structure, we rigorously prove LRE in $\ket{\Psi}$. We prove that $\ket{\Psi}$ has zero correlation length for arbitrary local operators; and $\ket{\Psi}$ cannot be disentangled through constant-depth local quantum circuits. More importantly, our proof does not rely on the code structure, but on the self-similarity in the entanglement pattern which is right the property that is believed incompatible with TO~\cite{Zhu2021}. This mark an intrinsic difference between $\ket{\Psi}$ and TO states, and suggests that the fractal self-similarity might underlie a new paradigm for the realization of LRE in quantum states.

\section{Tensor-network construction}
$\ket{\Psi}$ is a fixed point of wavefunction renormalization with a self-similar entanglement pattern, and is contracted from a ``self-similar'' tensor-network. The tensor network consists of copies of a single tensor $A=\sum_{\alpha\beta\beta'\beta''}A(\alpha\beta\beta'\beta'')\ket{\alpha\otimes\beta\beta'\beta''}\in\mathbb{C}^d\otimes\mathbb{C}^{D^3}$ with the physical index $\ket{\alpha}=\ket{0},\ket{1},\ket{2},\ket{3}\in\mathbb{C}^d$ for the local qudit ($d=4$), and three virtual indices $\ket{\beta},\ket{\beta'},\ket{\beta''}=\ket{0},\ket{1}\in\mathbb{C}^D$ ($D=2$) resembling the linking between neighboring qudits (see Fig.~\ref{1c}). At finite system size, the contraction is together with three constant corner tensors $C_1,C_2,C_3=\ket{1}\in\mathbb{C}^2$.

The characterization of the entanglement pattern of $\ket{\Psi}$ is based on the nonzero contraction of tensors, and the self-similarity is encoded in the tensor $A$ subject to a scale-invariance condition $W^+\Tr_{\{\mathbb{C}^D\}}[A\otimes A \otimes A]=\lambda A$. As illustrated in Fig.~\ref{1c}, the scale invariance equates the fixed-point condition with the self-similarity (zoom-in and zoom-out invariance) of the fractal: contracting the three copies of $A$ in each block followed by $W^+$, i.e. zooming out, results in $A$ itself times a constant $\lambda$, i.e. invariance. Here $W:\mathbb{C}^d\rightarrow\mathbb{C}^{d^3}$ is an isometry, and $W^+$ extracts the degrees of freedom inside a three-qudit block (see Fig.~\ref{1b}) in charge of its entanglement with neighboring blocks so that $(W^+)^\otimes$ plays the role of coarse graining.

As a solution to the scale-invariance condition, the tensor $A$ is defined as $A=A_0+A_1+A_2+A_3$ with $A_0=\ket{0}\otimes\ket{000+111}$, $A_1=\ket{1}\otimes\ket{100+011}$, $A_2=\ket{2}\otimes\ket{010+101}$ and $A_3=\ket{3}\otimes\ket{001+110}$. For convenience in illustration (see Fig.~\ref{1d}), we represent $000+111$ by ${1\atop{0}}{1\atop{0}}{1\atop{0}}$, $100+011$ by ${0\atop{1}}{1\atop{0}}{1\atop{0}}$, etc. In contraction, we take the convention that each tensor inside a block is outward (see Fig.~\ref{1e} and \ref{1f}). The details of the contraction is given in Appendix.

\section{Self-similar entanglement pattern}
\begin{figure}[ht]
\centering
    \includegraphics[width=8.3cm]{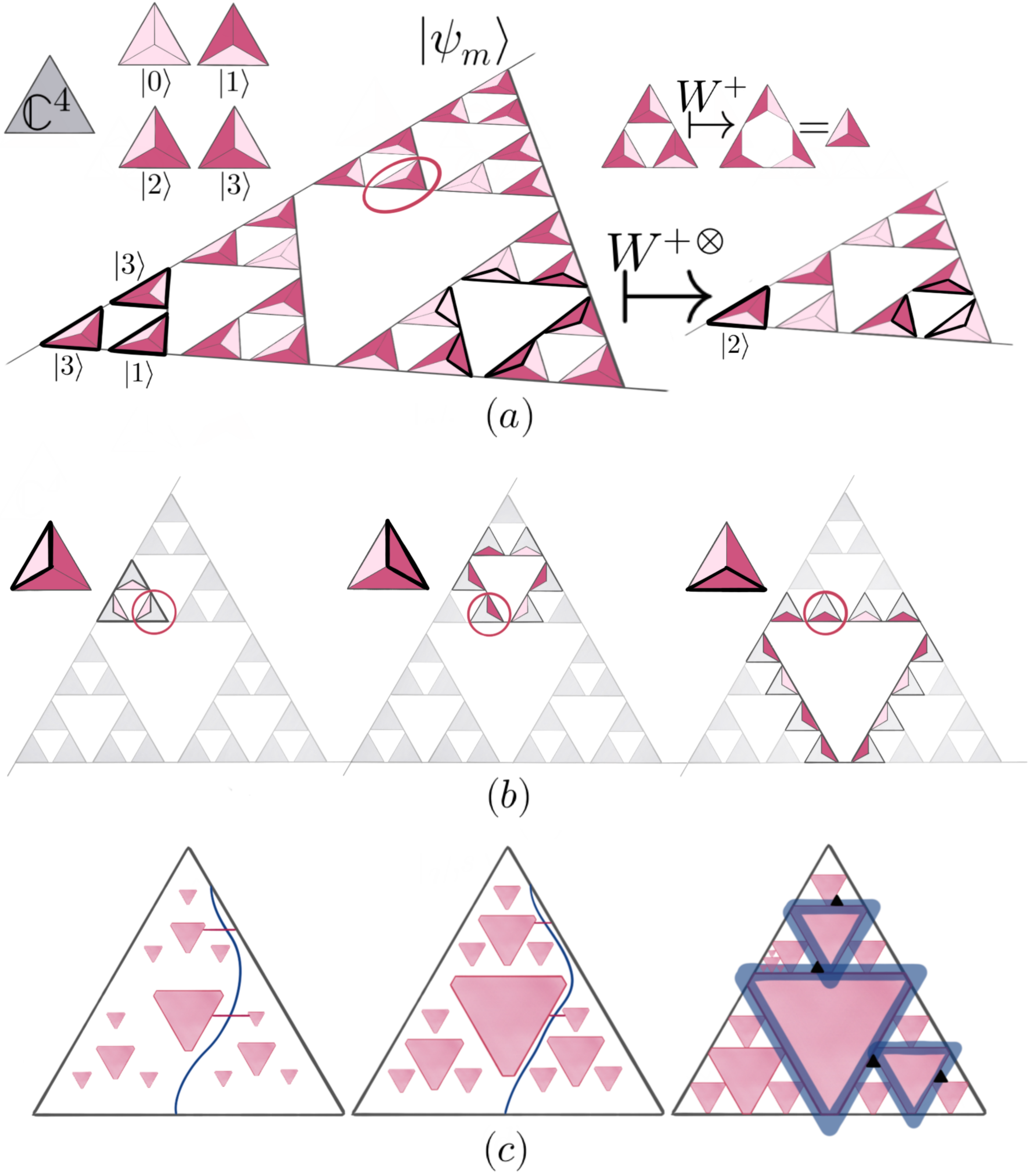}   
\phantomsubfloat{\label{2a}}\phantomsubfloat{\label{2b}}\phantomsubfloat{\label{2c}}
\vspace{-2\baselineskip}
\caption{(a) The constraints and scale invariance of $\ket{\psi_m}$. The colored-triangles representation of $\ket{\alpha}$ with the same convention of orientation as in the tensor contraction (see the highlighted triangles). (b) The loop structure of different sizes in the representation of $\ket{\psi_m}$. Each vertex (in the red circle) appears exactly in three loops. In (c) the left two are lattices with holes (the pink regions) where the white bulk represents vertices in a regular geometry with continuum limit. Keeping enlarging the size of holes, the left converges to the right on which keeping punching holes results in the Sierpi\'nski lattice. The bold lines in the right lattice illustrate the interlocking triangular loops.}
\end{figure}

The tensor network contracts to $\ket{\Psi}=1/\sqrt{M}\sum_m\ket{\psi_m}$, an equal-weight sum, with qudit-product state $\ket{\psi_m}=\ket{\alpha\alpha'\alpha''\cdots}=\Tr_{\{\mathbb{C}^2\}}[A_3\otimes A_2\otimes A_0 \cdots C_1\otimes C_2\otimes C_3]$. To characterize the entanglement pattern, we illustrate the constraints that specify $\ket{\psi_m}$. As shown in Fig.~\ref{2a}, we represent the single-qudit state $\ket{\alpha}$ by colored triangle with pink and red sides. Thus in the same convention of orientation, the triangles for $\ket{0}$, $\ket{1}$, $\ket{2}$ and $\ket{3}$ are consistent with $A_0, A_1, A_2$ and $A_3$ in the contraction (see Fig.~\ref{1d} and \ref{2a}): red sides correspond to the pair of virtual indices of a tensor with a flip between ${1\atop{0}}$ and ${0\atop{1}}$.

An important structure for pictorially characterizing the constraints is the triangular loops (see Fig.~\ref{2b}), or simply loops, as formed by the colored sides of the triangles ($\ket{\alpha}$) along each closed ring structure of vertices in the lattice such that each qudit (vertex) appears in three such loops. For convenience, we treat the three lattice laterals and the loops equally. And by ``loop'' ,we refer to the ring structure of both the colored sides and the underlying vertices when it is clear in the context.

Then, the constraints are associated to each loop and simply read: the number of red sides in each loop is even (see Fig.~\ref{2a} and \ref{2b}). The constraints guarantee the genuine multipartite entanglement of $\ket{\Psi}$, since any bipartition crosses loops.

These constraints have two properties intrinsic to fractality, which characterize the self-similar entanglement pattern and determine the LRE of $\ket{\Psi}$: (1) Each qudit is engaged in three constraints (on three loops) so that all qudits are entangled through the interlocking structure of loops in all different length scales (see Fig.~\ref{2b} and \ref{2c}). (2) The interlocking structure of constraints is the same in all length scales, i.e. zoom-in and zoom-out invariant. The invariance is illustrated in Fig.~\ref{2a} where the coarse graining ${W^+}^{\otimes}$ (see formal definition in Appendix) removes the inner sides of each block (the smallest loop) which are not included in constraints linking blocks (larger loops), and forms new colored triangles ($\ket{\alpha}$'s in the larger scale) by reducing the outer six sides remained in each block following the rule: two red (pink) sides are reduced into a pink side, and one red plus one pink side are reduced into a red side. Obviously, the coarse-grained state satisfies the same constraints.

\section{Incompatibility with TO}
Before proving LRE of $\ket{\Psi}$, we show how its self-similar entanglement pattern is incompatible with the code structure for TO using the following heuristic arguments. We consider a TO in a comparable setting, i.e. a quantum code with macroscopic code distance and described by homology on a porous 2D lattice illustrated in Fig.~\ref{2c} (the left two) where closed or open gapped boundaries around holes (the colored region) and the laterals condensate anyons (see the detail in Ref.~\cite{Delfosse2016,Zhu2021}). We show that the macroscopic code distance vanishes when turning the lattice geometry to our case.

Required by the TO structure, the bulk (the white region as the continuum limit) provides the background for deformable homologically nontrivial string-like logical operators, on which the shortest strings (the red line) and the longest strings (the blue line) of different types intersect for odd times and thus do not commute (see Fig.~\ref{2c}). Due to the non-commutativity, the shortest string confines the code distance and is hence required to increase with the system size in order to guarantee the nonlocal properties, e.g., the LRE of TO state. Consequently, in tuning the lattice geometry to our case by enlarging the holes (squeezing the the shortest distance between them) until the interlocking structure of loops emerges, the shortest string operator will be eventually squeezed into a local operator (see the right of Fig.~\ref{2c}) so that the code distance is no longer macroscopic. Hence the entanglement pattern of $\ket{\Psi}$ is excluded from TO.

\begin{figure}[!ht]
\centering
    \includegraphics[width=8.6cm]{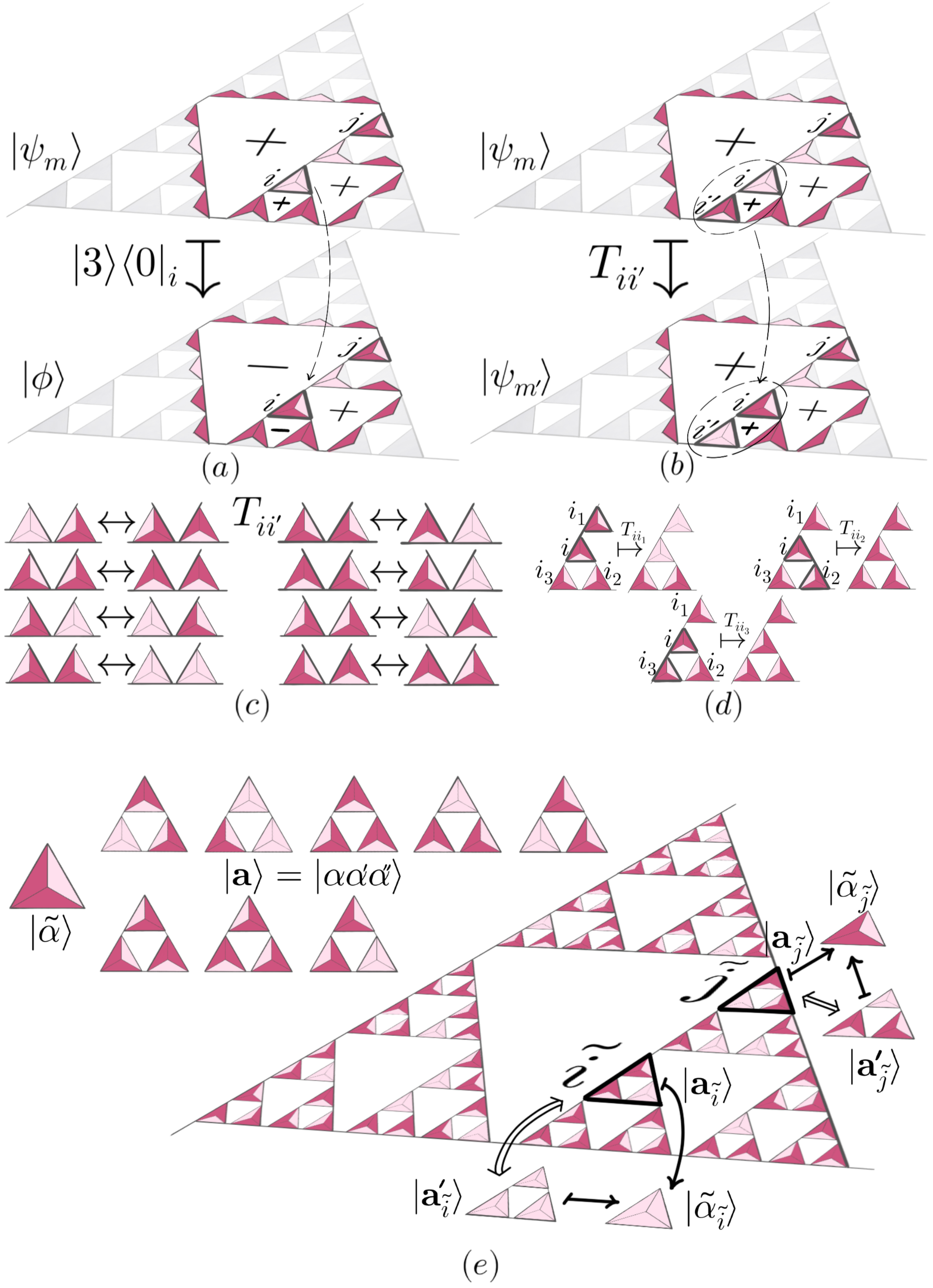}   
\phantomsubfloat{\label{3a}}\phantomsubfloat{\label{3b}}\phantomsubfloat{\label{3c}}\phantomsubfloat{\label{3d}}\phantomsubfloat{\label{3e}}
\vspace{-2\baselineskip}
\caption{(a) Operator $\dyad{3}{0}_i$ for the qudit on vertex $i$ maps $\ket{\psi_m}$ to $\ket{\phi}$. $``+''$ for constraints satisfied on the corresponding loops and $``-''$ for opposite constraint. (b) Operator $T_{ii'}$ maps $\ket{\psi_m}$ to $\ket{\psi_{m'}}$, preserving all constraints. (c) The content of $T_{ii'}$ acting on a pair of neighboring qudits. (d) $T_{ii_1},T_{ii_2},T_{ii_3}$ acting on different neighboring pairs of qudits map the single-qudit-state $\ket{1}$ on vertex $i$ to $\ket{0}$, $\ket{3}$ and $\ket{2}$ respectively. (e) The left: the eight three-qudit $\ket{\mathbf{a}}$ states (in a triangular block) that shares the coarse-grained state $\ket{\tilde{\alpha}}$. The right: local replacement of $\ket{\mathbf{a}_{\tilde i}}$ and $\ket{\mathbf{a}_{\tilde j}}$ on blocks $\tilde i$ and $\tilde j$ while keeping the coarse-grained $\ket{\tilde{\alpha}_{\tilde i}}$ and $\ket{\tilde{\alpha}_{\tilde j}}$.}
\end{figure}

\section{Proof for the zero correlation length of single-qudit operators} To prove the LRE of $\ket{\Psi}$, we firstly prove that $\ket{\Psi}$ has zero correlation length. In principle, we have to go through all local operators with arbitrary finite supports. However, thanks to the self-similar entanglement pattern, the proof can be considerably simplified.

We begin with proving for the case of single-qudit operators, i.e. $\expval{O_jO_i}{\Psi}-\expval{O_j}{\Psi}\expval{O_i}{\Psi}=0$ for any local operators $O_i$ and $O_j$ acting on nonadjacent qudits $i$ and $j$. We sketch the proof here. The complete proof is given in Appendix.

According to the expansion $O_i=\sum_{\alpha'\alpha}c^i_{\alpha'\alpha}\dyad{\alpha'}{\alpha}, O_j=\sum_{\alpha'\alpha}c^j_{\alpha'\alpha}\dyad{\alpha'}{\alpha}$, it suffices to show that $\expval{(\dyad{\bar{\alpha}'}{\bar{\alpha}}_j\dyad{\alpha'}{\alpha}_i)}{\Psi}-\expval{(\dyad{\bar{\alpha}'}{\bar{\alpha}}_j)}{\Psi}\expval{(\dyad{\alpha'}{\alpha}_i)}{\Psi}=0$ for any qudit basis states $\ket{\alpha},\ket{\alpha'},\ket{\bar{\alpha}},\ket{\bar{\alpha}'}$.

For the case where $\ket{\alpha}\ne\ket{\alpha'}$ or $\ket{\bar{\alpha}}\ne\ket{\bar{\alpha}'}$, the key in the proof is the fact that each qudit is engaged in constraints on three loops (including the laterals), and any two nonadjacent qudits share at most one constraints (see Fig.~\ref{2a}, \ref{2b} and \ref{3a}) so that qudit-product states $(\dyad{\alpha'}{\alpha}_i)\ket{\psi_m}$, $(\dyad{\bar{\alpha}}{\bar{\alpha}'})\ket{\psi_{m'}}$ and $\ket{\psi_{m''}}$ differ in at least one constraint from each other, and are hence orthogonal to each other (see Fig.~\ref{3a}). Then, it is easy to show the expectations in the correlation function are zero.

For the case where $\ket{\alpha}=\ket{\alpha'}$ and $\ket{\bar{\alpha}}=\ket{\bar{\alpha}'}$, we show that $\expval{(\dyad{\alpha}{\alpha}_i)}{\Psi}=\expval{\dyad{\bar{\alpha}}{\bar{\alpha}}_j}{\Psi}=1/4$ and $\expval{(\dyad{\bar{\alpha}}{\bar{\alpha}}_j\dyad{\alpha}{\alpha}_i)}{\Psi}=1/16$. The key is the existence of unitary operator $T_{ii'}$ (formally defined in Appendix) on neighboring qudits as pictorially defined in Fig.~\ref{3c}, which preserves the constraints and can map the state $\ket{\alpha}$ of a qudit on vertex $i$ to the other three basis states respectively with suitably chosen neighboring vertex $i'$ (see Fig.~\ref{3b} and \ref{3d}). Then, there are one-to-one mapping between $(\dyad{0}{0}_i)\ket{\psi_m}$'s, $(\dyad{1}{1}_i)\ket{\psi_{m'}}$'s, $(\dyad{2}{2}_i)\ket{\psi_{m''}}$'s and $(\dyad{3}{3}_i)\ket{\psi_{m''''}}$'s so that we can derive the desired results.

\section{Proof for the zero correlation length of operators with finite support} 
We now generalize the zero-correlation-length property and prove that $\expval{O_{\tilde j}O_{\tilde i}}{\Psi}-\expval{O_{\tilde j}}{\Psi}\expval{O_{\tilde i}}{\Psi}=0$ for local operators $O_{\tilde i}$ and $O_{\tilde j}$ supported on local finite regions ${\tilde i}$ and ${\tilde j}$ that are apart. Here, we sketch the proof for the case where $O_{\tilde i}$ and $O_{\tilde j}$ are supported on two blocks ${\tilde i}$ and ${\tilde j}$ and are separated by at least one block (see Fig.~\ref{3e}). In Appendix, we give the complete proof, and also show how the proof is directly generalized to the most general case with the same idea: Utilize the self-similar entanglement pattern to reduce the correlation in question to that in the single-qudit case.

The key for the reduction is that the degrees of freedom of each block, represented by the basis $\{\ket{\alpha\alpha'\alpha''}\}$, can be decomposed through a local unitary operator $U_{\tilde i}$ into ``relevant'' ones and ``irrelevant'' ones. For simplicity, we use the notation $\ket{\mathbf{a}_{\tilde i}}=\ket{\alpha\alpha'\alpha''}$ for the $4\times 8$ $\ket{\alpha\alpha'\alpha''}$ states satisfying the constraints on the smallest loop in the block (every eight such states share one coarse-grained single-qudit state $\ket{\tilde{\alpha}}$, see Fig.~\ref{3e}). The unitary operator satisfies $U_{\tilde i}\ket{\mathbf{a}_{\tilde i}}=\ket{\tilde{\alpha}_{\tilde i}}\otimes\ket{\gamma_{\tilde i}}$. Here, the single-qudit state $\ket{\tilde{\alpha}_{\tilde i}}=\sqrt{8}W^+\ket{\mathbf{a}_{\tilde i}}$ represents the relevant degrees of freedom in the smaller lattice as zoomed-out through coarse graining where $\tilde i$ labels a qudit (see Fig.~\ref{3e}); $\ket{\gamma_{\tilde i}}=\ket{1},\ket{2},\cdots,\ket{8}\in\mathbb{C}^8$ labels the irrelevant degrees of freedom, i.e. the eight $\ket{\mathbf{a}_{\tilde i}}$ states sharing the same coarse-grained $\ket{\tilde{\alpha}_{\tilde i}}$ (see Fig.~\ref{3e}).

Importantly, as illustrated in Fig.~\ref{3e}, for an arbitrary $\ket{\psi_m}$, we can replace $\ket{\mathbf{a}''_{\tilde i}}$ and $\ket{\mathbf{a}'''_{\tilde j}}$ on any two blocks $\tilde i$ and $\tilde j$ independently by any other $\ket{\mathbf{a}_{\tilde i}}$ and $\ket{\mathbf{a}_{\tilde j}}$ while keeping their coarse-grained single-qudit state $\ket{\tilde{\alpha}_{\tilde i}}$ and $\ket{\tilde{\alpha}_{\tilde j}}$. The result is another $\ket{\psi_{m'}}$ (all constraints satisfied) which shares the same coarse-grained qudit-prodcut state $\ket{\widetilde{\psi}_{\tilde m}}$ with $\ket{\psi_{m'}}$. Because of the independence in the local replacement, we can show that the contribution in the correlation function from the irrelevant degrees of freedom $\ket{\gamma_{\tilde i}}$ can be canceled out, and the relevant degrees of freedom give rise to the coarse-grained state $\ket{\widetilde{\Psi}}$ of $\ket{\Psi}$.

Then, using the relation $\expval{O_{\tilde j}O_{\tilde i}}{\Psi}-\expval{O_{\tilde j}}{\Psi}\expval{O_{\bar i}}{\Psi}=\expval{UO_{\tilde j}U^+UO_{\bar i}U^+}{U\Psi}-\expval{UO_{\tilde j}U^+}{U\Psi}\expval{UO_{\tilde i}U^+}{U\Psi}$ with $U=\otimes_{\tilde i}U_{\tilde i}$, the correlation can be reduced into the correlation of single-qudit operators in $\ket{\widetilde{\Psi}}$, which has been prove to be zero in the previous section. Hence the desired property can be proved.

\begin{figure*}[!ht]
\centering
    \includegraphics[width=15cm]{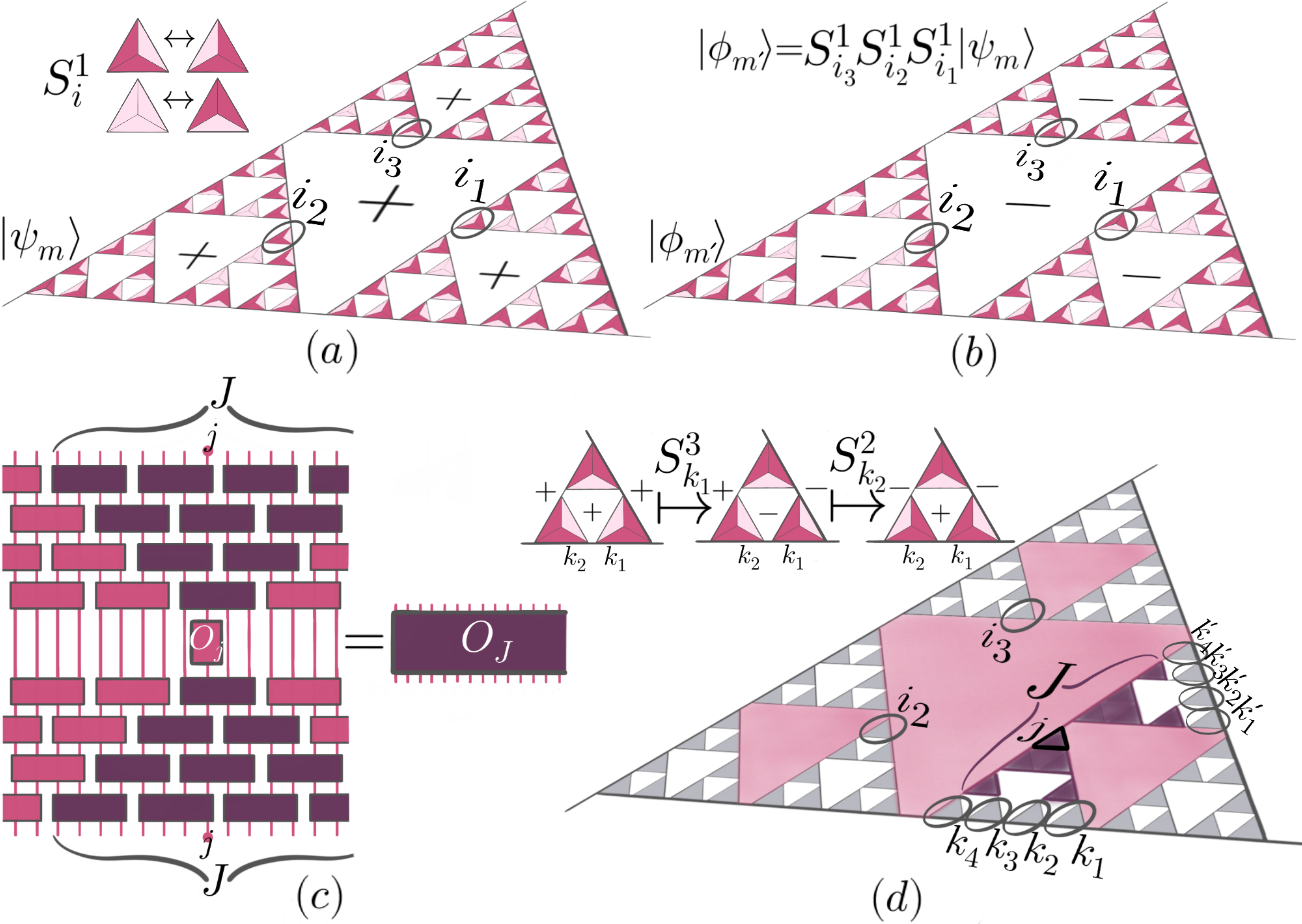}   
\phantomsubfloat{\label{4a}}\phantomsubfloat{\label{4b}}\phantomsubfloat{\label{4c}}\phantomsubfloat{\label{4d}}
\vspace{-0.5\baselineskip}
\caption{(a) The state $\ket{\psi_m}$ and the unitary operator $S^1_i$. (b) The state $\ket{\phi_{m'}}$. $``+''$ for constraint satisfied and $``-''$ for opposite constraint satisfied. $V_1=S^1_{i_3}S^1_{i_2}S^1_{i_1}$ maps $\ket{\psi_m}$ to $\ket{\phi_{m'}}$. (c) $U^+O_jU=O_J$ where $U=U_L\cdots U_2U_1$ is a local quantum circuits. The lines stand for qudits and the dark parts denote a causal structure based on which $U$ extends the support of $O_j$ as a single vertex to the support $J$ of $O_J$ as a connected region. (d) The support $J$ does not overlap with the path $K=\{k_1,k_2,k_3,k_4\}$ or $K'=\{k'_1,k'_2,k'_3,k'_4\}$. The unitary $V_2=S^1_{i_3}S^1_{i_2}S^2_{k_4}S^3_{k_3}S^2_{k_2}S^3_{k_1}$ or $V_3=S^1_{i_3}S^1_{i_2}S^3_{k'_4}S^2_{k'_3}S^3_{k'_2}S^2_{k'_1}$ maps $\ket{\psi_m}$ to $\ket{\phi_{m'}}$.}
\end{figure*}

\section{Proof for the system-size dependence of the circuit depth}
In the fractal geometry, the concepts of bulk and boundary do not have distinct meaning, and hence the argument of topological entanglement entropy does not directly apply. We need to directly apply the arguments of local quantum circuits. A local quantum circuit is a unitary operator $U=U_L\cdots U_2U_1$. Here, each $U_l=\otimes_pU_l(p)$ is a product of unitary operators acting on nonoverlapping local patches $\{p\}$ of qudits with sizes smaller than some number $P$, and $L$ is the depth of the circuit (see Fig.~\ref{4c}). We will prove that for arbitrarily given $L$ and $P$, any local quantum circuit characterized by $L$ and $P$ cannot completely disentangle $\ket{\Psi}$ when the system is over certain size. We will also give the dependence of the lower bounds of $L$ and $P$ on the system size for a local quantum circuit to completely disentangle $\ket{\Psi}$. We sketch the proof here. The complete proof is given in Appendix.

We introduce another state $\ket{\Phi}$ which is orthogonal to $\ket{\Psi}$. Here $\ket{\Phi}=1/\sqrt{M}\sum_m\ket{\phi_m}$ is the sum over all qudit-product-states satisfying all constraints as in $\ket{\psi_m}$ except on the largest four loops where the opposite constraints (odd number of red sides) are satisfied (see Fig.~\ref{4a} and \ref{4b}). In the proof, we utilize the property of $\ket{\Psi}$ and $\ket{\Phi}$ for detecting errors of the form $U^+O_jU$. Here $O_j$ is a local operator located at some vertex $j$ and $U$ dress it up so that $U^+O_jU$ has enlarged support $J$ (see Fig.~\ref{4c} and \ref{4d}). By definition, $\ket{\Psi}$ and $\ket{\Phi}$ detects an error $O$ if and only if $\mel{\Psi}{O}{\Phi}=\mel{\Phi}{O}{\Psi}=0$ and $\expval{O}{\Psi}=\expval{O}{\Phi}$~\cite{KitaevBook2002}.

Note that any subspace including $\ket{\Psi}$ and $\ket{\Phi}$ cannot form a quantum code with macroscopic code distance, because a unitary operator $V_1=S^1_{i_3}S^1_{i_2}S^1_{i_1}$ supported on three qubits (see Fig.~\ref{4b}) can simply map $\ket{\Psi}$ to $\ket{\Phi}$ and hence the code distance is limited by $3$. Pictorially (see Fig.~\ref{4a}), unitary operator $S^1_{i},S^2_{i}$ or $S^3_{i}$ supported on a qudit $i$ exchanges the red and pink on two sides of
the triangle or create two red (pink) sides from two pink
(red) sides. The superscript indicates the three choices of sides (see Appendix for formal definition).

Using the error-detecting property is inspired by the following consideration. If we assume the contrary condition that $U$ completely disentangles $\ket{\Psi}$, i.e. $U\ket{\Psi}=\otimes_i\ket{\chi_i}$ with $\ket{\chi_i}$ being a normalized single-qudit-state on vertex $i$, then, since $\braket{U\Phi}{U\Psi}=\mel{\Phi}{U^+U}{\Psi}=\braket{\Phi}{\Psi}=0$, it is easy to show that there exists a local operator $O_j$ located at some vertex $j$ such that $\expval{O_j}{U\Psi}\ne\expval{O_j}{U\Phi}$, i.e. $\expval{U^+O_jU}{\Psi}\ne\expval{U^+O_jU}{\Phi}$, and the errors $U^+O_jU$ cannot be detected.

This observation prompts us to consider proof by contradiction, i.e., we simply need to prove that $\ket{\Psi}$ and $\ket{\Phi}$ detect all errors of the form $O_J=U^+O_jU$ once the system is over certain size. The key for proving the property relies on the graph-theoretical property of the interlocking-loop geometry that when the system size is large enough, $O_J$ always commutes with a unitary $V$ and $V^+$ composed of operators like $S^1_{i},S^2_{i'},S^3_{i''}$, which maps $\ket{\Psi}$ to $\ket{\Phi}$ (see Fig.~\ref{4d}). Then, we can show $\expval{O_J}{\Phi}=\expval{O_J}{V\Psi}=\expval{V^+VO_J}{\Psi}=\expval{O_J}{\Psi}$.

\section{Discussion}
We have proved the LRE in $\ket{\Psi}$, i.e., the zero correlation length of $\ket{\Psi}$ and that $\ket{\Psi}$ cannot be disentangled through constant-depth local quantum circuits. To the best of our knowledge, $\ket{\Psi}$ is the first proved example of LRE state in fractional dimension between 1D and 2D. Our proof relies on the fractal self-similarity which is incompatible with the TO paradigm. The result might be unexpected, since LRE has been viewed as the microscopic equivalence of TO; it suggests that fractality, an intrinsic property in fractional spatial dimension, might underlie a novel paradigm for the realization of LRE in many-body quantum states. Indeed, since $\ket{\Psi}$ is defined from a single tensor, a class of new LRE states might be similarly constructed and further demonstrates the paradigm.

Strictly speaking, the current tools, e.g., the quantum circuit, cannot systematically characterize how nonlocal information is shared by local degrees of freedom and dissolves the correlation in LRE states. Hence, we cannot exclude the possibility that the entanglement in $\ket{\Psi}$ and that in some TO state have certain similarity. This uncertainty calls for a quantum-information framework that is beyond the specified locality and geometry. If within such a desired framework the LRE in $\ket{\Psi}$ is shown intrinsically different from that in TO state, a new equivalence between entanglement and quantum order can be systematically established. If instead certain intrinsic invariance can be proved, the LRE will be shown more profound than any characterization of quantum order in specified geometry or dimensions. Either way, our result paves the way for new discoveries connecting quantum information and quantum matter, and possibly application in relevant study of quantum computation.

We end the main text with preliminary discussion on how $\ket{\Psi}$ might be stabilized as ground state and how it can be possibly realized in experiments. As fractal quantum matter has become a new player in the discovery of exotic quantum emergent phenomena, it is intriguing to study the possible quantum order featuring the new LRE pattern. The comprehensive study on these topics requires another systematical work and will be included in our future research.

The new LRE pattern determines degenerate-ground-state property in studying local model stabilizing $\ket{\Psi}$: The state $\ket{\Phi}$ from the above proof is locally indistinguishable to $\ket{\Psi}$, hence it should be a degenerate ground state, which implies the absence of associated order parameter~\cite{Furukawa2006,Zeng2019}; however in this case, the code distance for the ground-state subspace, as shown in the last proof, is not macroscopic. This is another evidence that $\ket{\Psi}$ cannot be a ground state of TO but might rather, as a fixed point, portrays an exotic quantum order. The local indistinguishability with constant code distance might reflect the difference from TO in how nonlocal information is shared by qudits: Intuitively, the sharing in $\ket{\Psi}$ is hierarchically via the interlocking loops, and is ``cheaper'' than that in TO states where the sharing is uniform and gives rise to the braiding of local excitations (anyons). Accordingly, we expect exotic nonlocal properties of excitations from $\ket{\Psi}$.

A potential experimental approach for studying the peculiarity of the new LRE pattern without resorting to Hamiltonian is to directly prepare $\ket{\Psi}$ from $\ket{000\cdots}$ (embedded in 2D) through quantum circuits (on fractal geometry) in quantum processor where properties can be studied through measurements~\cite{Satzinger2021,Chi2022}. In principle, the feasibility is guaranteed by (1) ergodicity of the two-body local unitary operator $T_{ii'}$ (see the first proof) in the subspace spanned by $\{\ket{\phi_m}\}$ so that we have $\ket{\Psi}\propto\prod_{ii'}\frac{\mathds{1}+T_{ii'}}{\sqrt{2}}\ket{00\cdots}$; (2) circuit realization of $\{\mathds{1}+T_{ii'}\}$~\cite{Chi2022}. The proof of ergodicity is given in Appendix, it shows how local operators can stabilize the quantum fluctuation that generates the new LRE pattern. This also paves the way for searching potential realistic models stabilizing $\ket{\Psi}$.

\section*{Acknowledgements}
The author thanks Anne E. B. Nielsen for inspiring discussions on the tensor-network description of self-similarity and the long-range entanglement. The author thanks Wei Ku, Jinwu Ye and Jianda Wu for fruitful discussions on quantum order and quantum correlations. These discussions partially motivated this work. The author also thanks Chi Ming Yim, Haiyuan Zou, Fangyuan Gu, Zhen Wang, Weikang Lin and Yuzhu Cui for useful comments on the manuscript. This work is supported by the Tsung-Dao Lee Institute Postdoctoral Fellowship Grant.

\bibliographystyle{quantum}
\bibliography{ref2022}

\onecolumn


\appendix

\section{Contraction of the ``self-similar'' tensor network}
By definition, the contraction of copies of $A$ is a linear combination of contractions of copies from $A_0, A_1, A_2$ and $A_3$ so that we can focus on the latter. In contraction, we take the convention that each tensor inside a block is outward (see Fig.~1e in the main text). Then as shown in Fig.~1e and 1f in the main text, the contraction is nonzero if and only if there are even number of flips between ${1\atop{0}}$ and ${0\atop{1}}$ among the contracting indices, or there are even number of copies from $A_2$ or $A_3$ in any triangular block. Indeed, such non-vanishing contraction results in a big tensor of the same form, e.g., $A_3,A_2,A_0,\ldots$ contract to $\ket{320\ldots}\otimes\ket{{1\atop{0}}{1\atop{0}}{1\atop{0}}}$ as shown in Fig.~1f in the main text, which, further contracted with the constant corner tensors $C_1,C_2,C_3=\ket{1}\in\mathbb{C}^2$, results in a qudit-product state, e.g., $\ket{320\ldots}$. Note that the corner tensors only pick up big tensors of the form $\ket{\{\alpha\}}\otimes\ket{{1\atop{0}}{1\atop{0}}{1\atop{0}}}$ and ensures the equal-weight summation of $\ket{\Psi}$ as contracted from the tensor network: $\ket{\Psi}=1/\sqrt{M}\sum_m\ket{\psi_m}$ with $\ket{\psi_m}=\ket{\alpha\alpha'\alpha''\cdots}=\Tr_{\{\mathbb{C}^2\}}[A_3\otimes A_2\otimes A_0 \cdots C_1\otimes C_2\otimes C_3]$.

\section{Formal definition of $W^+$}
Formally, $W^+$ is defined as $1/\sqrt{8}$ times the sum
\begin{align*}
\begin{split} 
&\dyad{0}{000+111+023+132+213+230+302+321}\\
+&\dyad{1}{011+032+100+123+202+221+313+330}\\
+&\dyad{2}{010+033+101+122+203+220+312+331}\\
+&\dyad{3}{001+110+022+133+212+231+303+320}.
\end{split} 
\end{align*}

\section{Complete proof for the zero correlation length of single-qudit operators}
To complete the proof sketched in the main text, we first consider the case where $\ket{\alpha}\ne\ket{\alpha'}$ (the case of $\ket{\bar{\alpha}}\ne\ket{\bar{\alpha}'}$ can be treated in the same way). We observe that in the illustration of an arbitrary $\ket{\psi_m}$ (see Fig.~3c, 3d and 3a in the main text), each triangle (a qudit basis state) is engaged in constraints on three loops (including the laterals) respectively as each triangle has three sides. It follows that the change from $\ket{\alpha}$ to $\ket{\alpha'}$ (as the color change on two sides of the triangle) on a vertex $i$ will break the constraints on two loops (see Fig.~3a in the main text), resulting in a qudit-product-state $\ket{\phi}$ with opposite constraints (odd number of red sides) on the two loops and hence orthogonal to all the qudit-product-states expanding $\ket{\Psi}$. Then, if $\ket{\bar{\alpha}}=\ket{\bar{\alpha}'}$, we have $(\dyad{\bar{\alpha}}{\bar{\alpha}}_j)\ket{\psi_{m'}}$ (either unchanged or equal to $0$) is orthogonal to $\ket{\phi}=(\dyad{\alpha'}{\alpha}_i)\ket{\psi_m}$ and hence both $\expval{(\dyad{\bar{\alpha}}{\bar{\alpha}}_j\dyad{\alpha'}{\alpha}_i)}{\Psi}$ and $\expval{(\dyad{\bar{\alpha}}{\bar{\alpha}}_j)}{\Psi}\expval{(\dyad{\alpha'}{\alpha}_i)}{\Psi}$ are zero. On the other hand, if $\ket{\bar{\alpha}}\ne\ket{\bar{\alpha}'}$, we also observe that the vertex $j$ which is nonadjacent to $i$ share at most one loop with $i$ and the qudits on the two vertices respectively are at most commonly engaged in one constraint. Consequently, the change from $\ket{\bar{\alpha}'}$ to $\ket{\bar{\alpha}}$ on $j$ maps $\ket{\psi_{m'}}$ to another qudit-product-state $\ket{\phi'}$ which is orthogonal to both $\ket{\psi_m}$ and $\ket{\phi}$ (as they differ by at least the constraint on one loop). Hence, both $\expval{(\dyad{\bar{\alpha}'}{\bar{\alpha}}_j\dyad{\alpha'}{\alpha}_i)}{\Psi}$ and $\expval{(\dyad{\bar{\alpha}'}{\bar{\alpha}}_j)}{\Psi}\expval{(\dyad{\alpha'}{\alpha}_i)}{\Psi}$ are zero, and we have the desired condition.

Then we consider the other case where $\ket{\alpha}=\ket{\alpha'}$ and $\ket{\bar{\alpha}}=\ket{\bar{\alpha}'}$. We firstly show that $\expval{(\dyad{\alpha}{\alpha}_i)}{\Psi}=\expval{\dyad{\bar{\alpha}}{\bar{\alpha}}_j}{\Psi}=1/4$ for any basis state $\ket{\alpha},\ket{\bar{\alpha}}$. To that end, we define local unitary operator $T_{ii'}$ for any pair of qudits on neighboring vertices $i$ and $i'$ as specified in Fig.~3c in the main text. Note that $T_{ii'}$ has different forms for $(i,i')$ in the same block and in two neighboring blocks using the operator defined in the subsection for the quantum circuit, i.e., $T_{ii'}=S^1_iS^1_{i'}$ for $(i,i')$ connecting two neighboring blocks and $T_{ii'}=S^2_iS^3_{i'}$ or $T_{ii'}=S^3_iS^2_{i'}$ for $(i,i')$ in the same block. However, the pictorial representation of $T_{ii'}$ in different cases is the same as given by Fig.~3c in the main text, thanks to the colored-triangle representation of the single-qudit-states. Three important properties of $T_{ii'}$ acting on $\ket{\psi_m}$ can be directly read from Fig.~3c in the main text: (1) $T_{ii'}$ preserves the constraints on all loops (including the laterals); (2) for a qudit on vertex $i$ and in any $\ket{\alpha}$, the three unitary operators $T_{ii_1},T_{ii_2},T_{ii_3}$ ($i_1,i_2,i_3$ the three neighboring vertices of $i$) maps $\ket{\alpha}$ exactly to the other three basis states respectively (see Fig.~3d in the main text); (3) For cases where $(i,i_2,i_3)$ forming a block while $i_1$ outside, $T_{ii_2}T_{ii_3}$ and $T_{ii_1}$ map the single-qudit-state on $i$ to the same state.

Now, as shown in Fig.~3b in the main text, if in $\ket{\psi_m}$ the qudit on vertex $i$ is in $\ket{\alpha}$, certain $T_{ii'}$ (or $T_{ii_2}T_{ii_3}$ for $i$ at the corner) maps $\ket{\psi_m}$ exactly to a $\ket{\psi_{m'}}$ with the single-qudit-state on $i$ changed from $\ket{\alpha}$ to $\ket{\alpha'}$. Since $T_{ii'}$ is unitary, it maps all $\ket{\psi_m}$'s with nonzero $\expval{(\dyad{\alpha}{\alpha}_i)}{\psi_m}$ one-to-one to all $\ket{\psi_{m'}}$'s with nonzero $\expval{(\dyad{\alpha'}{\alpha'}_i)}{\psi_{m'}}$. Therefore, since $\sum_{\alpha=0}^3\expval{(\dyad{\alpha}{\alpha}_i)}{\Psi}=1/M\sum_{\alpha=0}^3\sum_m\expval{(\dyad{\alpha}{\alpha}_i)}{\psi_m}=1$, we have$\expval{(\dyad{\alpha}{\alpha}_i)}{\Psi}=1/4$ and similarly $\expval{\dyad{\bar{\alpha}}{\bar{\alpha}}_j}{\Psi}=1/4$. Next, we show $\expval{(\dyad{\bar{\alpha}}{\bar{\alpha}}_j\dyad{\alpha}{\alpha}_i)}{\Psi}=1/16$. Indeed, following the spirit of the above arguments, we only need to notice that vertices $i$ and $j$ are nonadjacent so that $T_{ii'}$ leave the qudit on $j$ unchanged (see Fig.~3b in the main text), and hence we have $\expval{(\dyad{\bar{\alpha}}{\bar{\alpha}}_j\dyad{\alpha}{\alpha}_i)}{\Psi}=\expval{(\dyad{\bar{\alpha}}{\bar{\alpha}}_j\dyad{\alpha'}{\alpha'}_i)}{\Psi}=1/4\times 1/4=1/16$. Obviously, it means that $\expval{(\dyad{\bar{\alpha}}{\bar{\alpha}}_j\dyad{\alpha}{\alpha}_i)}{\Psi}-\expval{(\dyad{\bar{\alpha}}{\bar{\alpha}}_j)}{\Psi}\expval{(\dyad{\alpha}{\alpha}_i)}{\Psi}=1/16-1/16=0$. We have hence proved that $\ket{\Psi}$ has zero correlation length.

\section{Complete proof for the zero-correlation-length of operators with finite support}
To complete the proof sketched in the main text, we consider local operators $O_{\tilde i}$ and $O_{\tilde j}$ supported on two nonadjacent blocks ${\tilde i}$ and ${\tilde j}$ as illustrated in Fig.~3e in the main text. Before computing their correlation we study the degrees of freedom of each block, i.e. the space $(\mathbb{C}^4)^{\otimes 3}$ of the three qudits spanned by the basis $\{\ket{\alpha\alpha'\alpha''}\}$. According to the definition of $W^+$ (see Fig.~2a in the main text and above formal definition), there are $4\times 8$ $\ket{\alpha\alpha'\alpha''}$ states satisfying the constraints on the smallest loop in the block (every eight such states share one coarse-grained single-qudit state $\ket{\tilde{\alpha}}$, see Fig.~3e in the main text). For simplicity, we use the notation $\ket{\mathbf{a}_{\tilde i}}=\ket{\alpha\alpha'\alpha''}$ for these states (see Fig.~3e in the main text) and use $\ket{\mathbf{b}_{\tilde i}}$ for the rest $4\times 8$ states, and thus we can write $\ket{\psi_m}$ as $\ket{\mathbf{a}_{\tilde 1}\mathbf{a}'_{\tilde 2}\cdots\mathbf{a}''_{\tilde i}\cdots}$. Then, similar to the previous proof, since the operator $O_{\tilde i}$ and $O_{\tilde j}$ can be expanded on a basis formed by operators of the form $\dyad{\mathbf{a}}{\mathbf{a}'}$, $\dyad{\mathbf{b}}{\mathbf{b}'}$, $\dyad{\mathbf{a}}{\mathbf{b}}$ and $\dyad{\mathbf{b}}{\mathbf{a}}$, we only need to consider the correlation between those operators in the basis. However, due to the obvious fact that $(\dyad{\mathbf{a}_{\tilde i}}{\mathbf{b}_{\tilde i}})\ket{\psi_m}=(\dyad{\mathbf{b}_{\tilde i}}{\mathbf{b}_{\tilde i}})\ket{\psi_m}=0$ and $\bra{\psi_m}(\dyad{\mathbf{b}_{\tilde i}}{\mathbf{a}_{\tilde i}})=\bra{\psi_m}(\dyad{\mathbf{b}_{\tilde i}}{\mathbf{b}_{\tilde i}})=0$ (also for operators on $\tilde j$), we only need to consider $O_{\tilde i}$ and $O_{\tilde j}$ expanded on the $\dyad{\mathbf{a}}{\mathbf{a}'}$ operators.

We can further simplify the computation of the correlation by decomposing a three-qudit block into the tensor product of two parts as motivated by the following facts. As illustrated in Fig.~3e in the main text, for an arbitrary $\ket{\psi_m}=\ket{\mathbf{a}_{\tilde 1}\mathbf{a}'_{\tilde 2}\cdots\mathbf{a}''_{\tilde i}\cdots}$, we can replace $\ket{\mathbf{a}''_{\tilde i}}$ and $\ket{\mathbf{a}'''_{\tilde j}}$ on any two blocks $\tilde i$ and $\tilde j$ independently by any other $\ket{\mathbf{a}_{\tilde i}}$ and $\ket{\mathbf{a}_{\tilde j}}$ with $W^+\ket{\mathbf{a}_{\tilde i}}=W^+\ket{\mathbf{a}''_{\tilde i}}=1/\sqrt{8}\ket{\tilde{\alpha}_{\tilde i}}$ and $W^+\ket{\mathbf{a}_{\tilde j}}=W^+\ket{\mathbf{a}'''_{\tilde j}}=1/\sqrt{8}\ket{\tilde{\alpha}_{\tilde j}}$. The result is another $\ket{\psi_{m'}}$ (all constraints satisfied) with $(W^+)^\otimes\ket{\psi_{m'}}=(W^+)^\otimes\ket{\psi_m}=(1/\sqrt{8})^{N_b}\ket{\widetilde{\psi}_{\tilde m}}$ ($N_b$ the total number of blocks). Note that $\ket{\widetilde{\psi}_{\tilde m}}$ is simply the states satisfying all constraints on a smaller-size Sierpi\'nski lattice where $\tilde i$ labels a qudit (see Fig.~3e in the main text). Additionally, it can be easily checked that any $\ket{\widetilde{\psi}_{\tilde m}}$ is the coarse-grained state of some $\ket{\psi_m}$ since such a $\ket{\psi_m}$ can be simply constructed by replacing each $\ket{\tilde{\alpha}_{\tilde i}}$ with an arbitrary $\ket{\mathbf{a}_{\tilde i}}$ with $W^+\ket{\mathbf{a}_{\tilde i}}=1/\sqrt{8}\ket{\tilde{\alpha}_{\tilde i}}$ which will definitely keep all constraints satisfied in the zoomed-in lattice (see Fig.~3e in the main text). These facts implies that the degrees of freedom for choosing $\ket{\mathbf{a}_{\tilde i}}$ on each block can be separated from the coarse-grained ``skeleton'' $\ket{\widetilde{\psi}_{\tilde m}}$.

Based on the above facts, we can define a unitary operator $U=\otimes_{\tilde i}U_{\tilde i}$ with each $U_{\tilde i}:(\mathbb{C}^4)^{\otimes 3}\rightarrow\mathbb{C}^8\otimes\mathbb{C}^8$ supported on a three-qudit block $\tilde i$ and satisfying $U_{\tilde i}\ket{{\mathbf{a}}}=\ket{\tilde{\alpha}}\otimes\ket{\gamma}$. Here, $\ket{\tilde{\alpha}_{\tilde i}}=\sqrt{8}W^+\ket{{\mathbf{a}}}$ is the coarse-grained state, and $\ket{\gamma}=\ket{1},\ket{2},\cdots,\ket{8}\in\mathbb{C}^8$ labels $\ket{{\mathbf{a}}}$ among the eight states sharing the same coarse-grained $\ket{\tilde{\alpha}}$. We view the four $\ket{\tilde{\alpha}}$ states as embedded in $\mathbb{C}^8$ so that the $U_{\tilde i}\ket{{\mathbf{b}}}$ states span the orthogonal complement of the subspace spanned by the $\ket{\tilde{\alpha}}\otimes\ket{\gamma}$ states, i.e. $(\mathbb{C}^4\otimes\mathbb{C}^8)\oplus(\mathbb{C}^4\otimes\mathbb{C}^8)=\mathbb{C}^8\otimes\mathbb{C}^8$. Note that the arbitrariness in defining $U_{\tilde i}\ket{{\mathbf{b}}}$ and in the order of labeling the $\ket{{\mathbf{a}}}$ states will not hurt our arguments.

Now, we simply want to prove $\expval{O_{\tilde j}O_{\tilde i}}{\Psi}-\expval{O_{\tilde j}}{\Psi}\expval{O_{\bar i}}{\Psi}=\expval{UO_{\tilde j}U^+UO_{\bar i}U^+}{U\Psi}-\expval{UO_{\tilde j}U^+}{U\Psi}\expval{UO_{\tilde i}U^+}{U\Psi}=0$. And according to the sketch in the main text, we show how this correlation can be reduced to what we have proved for nonadjacent single-qudit operators, and hence the desired property can be proved.

According to the definition, we have $U\ket{\psi_m}=\otimes_{\tilde i}(U_{\tilde i}\ket{\mathbf{a}_{\tilde i}})=\otimes_{\tilde i}(\ket{\tilde{\alpha}_{\tilde i}}\otimes\ket{\gamma_{\tilde i}})=(\otimes_{\tilde i}\ket{\tilde{\alpha}_{\tilde i}})\otimes(\otimes_{\tilde i}\ket{\gamma_{\tilde i}})$. In the last step we identify the two Hilbert spaces $\otimes_{\tilde i}(\mathbb{C}^4\otimes\mathbb{C}^8)$ and $(\otimes_{\tilde i}\mathbb{C}^4)\otimes(\otimes_{\tilde i}\mathbb{C}^8)$. Furthermore, by definition, $\otimes_{\tilde i}\ket{\tilde{\alpha}_{\tilde i}}$ equals the coarse-grained state $(\sqrt{8})^{N_b}(W^+)^\otimes\ket{\psi_m}=\ket{\widetilde{\psi}_{\tilde m}}$, hence we have $U\ket{\psi_m}=\ket{\widetilde{\psi}_{\tilde m}}\otimes(\otimes_{\tilde i}\ket{\gamma_{\tilde i}})$. Then, consider $U\ket{\Psi}=1/\sqrt{M}\sum_m U\ket{\psi_m}$. Obviously, in the sum, $U\ket{\psi_m}=\ket{\widetilde{\psi}_{\tilde m}}\otimes(\otimes_{\tilde i}\ket{\gamma_{\tilde i}})$ goes through all the coarse-grained state $\ket{\widetilde{\psi}_{\tilde m}}$. Additionally, based on the above arguments about how different $\ket{\psi_m}$ states share the same coarse-grained state, we can conclude that for each $\ket{\widetilde{\psi}_{\tilde m}}$, the $\ket{\psi_m}$ states that mapped to it by $(W^+)^{\otimes}$ goes through all the possible $\otimes_{\tilde i}\ket{\gamma_{\tilde i}}$ states. Hence the $U\ket{\psi_m}$ states corresponding to the same $\ket{\widetilde{\psi}_{\tilde m}}$ sum up to 
\begin{equation*}
\sqrt{8}^{N_b}\sum_{\{\gamma_{\tilde i}\}}\ket{\widetilde{\psi}_{\tilde m}}\otimes(\otimes_{\tilde i}\ket{\gamma_{\tilde i}})=\sqrt{8}^{N_b}\ket{\widetilde{\psi}_{\tilde m}}\otimes(\otimes_{\tilde i}\ket{\Gamma_{\tilde i}})
\end{equation*}
with $\ket{\Gamma_{\tilde i}}=1/\sqrt{8}(\ket{1}+\ket{2}+\cdots+\ket{8})$ independent on $\tilde i$ and $\tilde m$. Then, we eventually have $U\ket{\Psi}=(\sqrt{8}^{N_b}/\sqrt{M})(\sum_{\tilde m}\ket{\widetilde{\psi}_{\tilde m}})\otimes(\otimes_{\tilde i}\ket{\Gamma_{\tilde i}})=\ket{\widetilde{\Psi}}\otimes(\otimes_{\tilde i}\ket{\Gamma_{\tilde i}})$ where $\ket{\widetilde{\Psi}}$ is the coarse-grained state of $\ket{\Psi}$, the state we have studied throughout the text but in the zoomed-out lattice with smaller size.

Note that $U$ does not disentangle the qudits in $\ket{\Psi}$, but instead represent $\ket{\Psi}$ in a different tensor product structure which  separates the degrees of freedom represented by the $\ket{\Gamma_{\tilde i}}$ states from the ``skeleton structure'' in the coarse-grained state $\ket{\widetilde{\Psi}}$. The ``skeleton structure'' entangles the degrees of freedom of blocks represented by $\ket{\tilde{\alpha}_{\tilde i}}$ and prevent the qudits from being disentangled by the local unitary operator $U$. The next section studies why even local quantum circuits cannot disentangle $\ket{\Psi}$.

Before computing the correlation, we recall that we only consider $O_{\tilde i}$ and $O_{\tilde j}$ that are expanded on operators of the form $\dyad{\mathbf{a}}{\mathbf{a}'}$. Since $U_{\tilde i}\dyad{\mathbf{a}_{\tilde i}}{\mathbf{a}'_{\tilde i}}U_{\tilde i}^+=\dyad{\tilde{\alpha}_{\tilde i}\otimes\gamma_{\tilde i}}{\tilde{\alpha}'_{\tilde i}\otimes\gamma'_{\tilde i}}$, $U_{\tilde i}O_{\tilde i}U_{\tilde i}^+$ is an operator on the subspace spanned by the $\ket{\tilde{\alpha}}\otimes\ket{\gamma}$ states, i.e. $\mathbb{C}^4\otimes\mathbb{C}^8\subset\mathbb{C}^8\otimes\mathbb{C}^8$ (and so is $U_{\tilde i}O_{\tilde j}U_{\tilde i}^+$). Then, the two operators can be expanded on an operator basis $\{o_{\kappa}\otimes q_{\kappa'}\}$ as $\sum_{\kappa \kappa'}c^{\tilde i}o^{\tilde i}_{\kappa}\otimes q^{\tilde i}_{\kappa'}$ and $\sum_{\kappa \kappa'}c^{\tilde j}o^{\tilde j}_{\kappa}\otimes q^{\tilde j}_{\kappa'}$, where $o_{\kappa}$ acts on $\ket{\tilde{\alpha}_{\tilde i}}\in\mathbb{C}^4$ and hence on $\ket{\widetilde{\Psi}}\in\otimes_{\tilde i}\mathbb{C}^4$, while $q_{\kappa'}$ acting on $\ket{\gamma_{\tilde i}}\in\mathbb{C}^8$ and hence on $\otimes_{\tilde i}\ket{\Gamma_{\tilde i}}\in\otimes_{\tilde i}\mathbb{C}^8$. Then, similar to the proof in the previous section, to show that the correlation is zero we only need to show $\expval{(o^{\tilde j}_{\kappa''}\otimes q^{\tilde j}_{\kappa'''})(o^{\tilde i}_{\kappa}\otimes q^{\tilde i}_{\kappa'})}{U\Psi}-\expval{o^{\tilde j}_{\kappa''}\otimes q^{\tilde j}_{\kappa'''}}{U\Psi}\expval{o^{\tilde i}_{\kappa}\otimes q^{\tilde i}_{\kappa'}}{U\Psi}=0$. Now, considering $U\ket{\Psi}=\ket{\widetilde{\Psi}}\otimes(\otimes_{\tilde i}\ket{\Gamma_{\tilde i}})$, we have 
\begin{multline*}
\expval{(o^{\tilde j}_{\kappa''}\otimes q^{\tilde j}_{\kappa'''})(o^{\tilde i}_{\kappa}\otimes q^{\tilde i}_{\kappa'})}{U\Psi}\\
=\expval{(o^{\tilde j}_{\kappa''}o^{\tilde i}_{\kappa})\otimes (q^{\tilde j}_{\kappa'''}q^{\tilde i}_{\kappa'})}{U\Psi}\\
=\expval{o^{\tilde j}_{\kappa''}o^{\tilde i}_{\kappa}}{\widetilde{\Psi}}\expval{q^{\tilde j}_{\kappa'''}q^{\tilde i}_{\kappa'}}{\otimes_{\tilde i'}\Gamma_{\tilde i'}}
\end{multline*}
and
\begin{multline*}
\expval{o^{\tilde j}_{\kappa''}\otimes q^{\tilde j}_{\kappa'''}}{U\Psi}\expval{o^{\tilde i}_{\kappa}\otimes q^{\tilde i}_{\kappa'}}{U\Psi}\\
=\expval{o^{\tilde j}_{\kappa''}}{\widetilde{\Psi}}\expval{q^{\tilde j}_{\kappa'''}}{\otimes_{\tilde i'}\Gamma_{\tilde i'}}\\
\times\expval{o^{\tilde i}_{\kappa}}{\widetilde{\Psi}}\expval{q^{\tilde i}_{\kappa'}}{\otimes_{\tilde i'}\Gamma_{\tilde i'}}.
\end{multline*}
Since for product state $\ket{\otimes_{\tilde i'}\Gamma_{\tilde i'}}$ we have $\expval{q^{\tilde j}_{\kappa'''}q^{\tilde i}_{\kappa'}}{\otimes_{\tilde i'}\Gamma_{\tilde i'}}=\expval{q^{\tilde j}_{\kappa'''}}{\Gamma_{\tilde j}}\expval{q^{\tilde i}_{\kappa'}}{\Gamma_{\tilde i}}=\expval{q^{\tilde j}_{\kappa'''}}{\otimes_{\tilde i'}\Gamma_{\tilde i'}}\expval{q^{\tilde i}_{\kappa'}}{\otimes_{\tilde i'}\Gamma_{\tilde i'}}$, it suffices to show $\expval{o^{\tilde j}_{\kappa''}o^{\tilde i}_{\kappa}}{\widetilde{\Psi}}-\expval{o^{\tilde j}_{\kappa''}}{\widetilde{\Psi}}\expval{o^{\tilde i}_{\kappa}}{\widetilde{\Psi}}=0$, which is exactly what we have studied in the previous section. Note that the fact that $O_{\tilde i}$ and $O_{\tilde j}$ are supported on nonadjacent blocks guarantees that $o^{\tilde i}_{\kappa}$ and $o^{\tilde j}_{\kappa''}$ are supported on nonadjacent qudits in the smaller-size lattice, and hence the correlation is zero. Therefore, we can conclude that $\expval{O_{\tilde j}O_{\tilde i}}{\Psi}-\expval{O_{\tilde j}}{\Psi}\expval{O_{\tilde i}}{\Psi}=0$.

For the general case where $O_{\tilde i}$ and $O_{\tilde j}$ are supported on two local regions of finite qudits, each of the two support can always be viewed as within a (larger) triangular block of certain number of qudits. Then, if we use $\tilde i$ and $\tilde j$ to denote the two blocks and just require them to be nonadjacent (separated by at least one such block), the above arguments can be directly applied except that multiple coarse-graining operators should be applied until the two blocks $\tilde i$ and $\tilde j$ are shrunk to two nonadjacent vertices. This eventually proves the zero-correlation-length property of $\ket{\Psi}$.

\section{Complete proof for the system-size dependence of the circuit depth}
We consider arbitrary $L$ and $P$ and an arbitrary local quantum circuit $U$ characterized by $L$ and $P$. We want to prove that once the system is over certain size, $U$ cannot completely disentangle $\ket{\Psi}$. Following the sketch in the main text where the state $\ket{\Phi}$ is introduced, we firstly assume the contrary condition and then use the arguments of the error-detecting property of $\ket{\Psi}$ and $\ket{\Phi}$ to show that the assumption implies a contradiction.

We start with assuming that $U$ completely disentangles $\ket{\Psi}$, i.e. $U\ket{\Psi}=\otimes_i\ket{\chi_i}$ with $\ket{\chi_i}$ being a normalized single-qudit-state on vertex $i$. Then, since $\braket{U\Phi}{U\Psi}=\mel{\Phi}{U^+U}{\Psi}=\braket{\Phi}{\Psi}=0$, it is easy to show that there exists a local operator $O_j$ located at some vertex $j$ such that $\expval{O_j}{U\Psi}\ne\expval{O_j}{U\Phi}$. Indeed, expanding $\ket{U\Phi}$ in an arbitrary qudit-product-state basis including $\otimes_i\ket{\chi_i}$, qudit-product-states contributing to the expansion are orthogonal to $\otimes_i\ket{\chi_i}$, and hence there must be some $\otimes_i\ket{\chi'_i}$ with $\braket{\chi'_j}{\chi_j}=0$ at some vertex $j$. In that case, we can take $O_j$ as $\mathds{1}\otimes\cdots\otimes\dyad{\chi_j}{\chi_j}\otimes\cdots\otimes\mathds{1}$ so that we have $\expval{O_j}{U\Phi}<1$ and hence $\expval{O_j}{U\Phi}\ne\expval{O_j}{U\Psi}$.

The inequality rewritten as $\expval{U^+O_jU}{\Phi}\ne\expval{U^+O_jU}{\Psi}$ implies that $\ket{\Psi}$ and $\ket{\Phi}$ cannot detect the error $U^+O_jU$. Then, to reach a contradiction, in the rest we prove that $\ket{\Psi}$ and $\ket{\Phi}$ detect all errors of the form $U^+O_jU$ once the system is over certain size.

Following the above consideration, we now prove that once the linear size $\mathcal{D}$ of the lattice is over $\frac{16}{3}PL-\frac{8}{3}P+\frac{5}{3}$ the error-detecting property is satisfied. To be consistent with the study of LRE in 2D topologically ordered states~\cite{Bravyi2006}, we take advantage of the graph theory and define the linear size of the lattice (a connected subset) as the diameter of the lattice (subset) in the graph (induced subgraph) metric. In the graph-theoretical language, in a connected graph, the distance between two vertices is the length of, i.e. the number of edges (links) in the shortest path connecting the two vertices. The diameter of a connected subset in a graph is the diameter of the subgraph induced by the subset, i.e., the maximal length of the shortest path within the subset which connects a pair of vertices in the subset. Then, $\mathcal{D}$ of a finite-generation Sierpi\'nski lattice is exactly the number of links on one lateral of the lattice as a triangle.

There is another important size in our proof which is determined by the causal structure in the local quantum circuit (see the dark color in Fig.~4c in the main text): the operator $O_J=U^+O_jU$ is supported on a connected subset $J$ of vertices with size $\mathcal{D}_J\le(2L-1)P$. Note that to be consistent with the case in 2D, a local patch of unitary operator of size $P$ in the quantum circuit is defined as a unitary operator supported on a connected subset with diameter $P$, i.e. it acts as the identity operator on qudits outside the support. The $P$ here confines the distance between any pair of vertices within the subset so that the unitary operator acts locally. Then, starting from a single vertex $j$ as the support of $O_j$, the first layer of non-overlapping local unitary operators extends the size of the support at most by $P$ since only one local patch contains $j$. And each of the following layers extends the support by $2P$ since the distance of the farthest apart vertex pair is extended at most by $2P$. Note that the layer structure in the quantum circuit guarantees the connectivity of $J$.

Now, we assume that the system is over certain size, i.e. the inequality $\mathcal{D}\ge\frac{16}{3}PL-\frac{8}{3}P+\frac{5}{3}$ or equivalently $\mathcal{D}_J\le\frac{3}{8}\mathcal{D}-\frac{5}{8}$, and prove the error-detecting property.

We firstly prove $\mel{\Psi}{O_J}{\Phi}=\mel{\Phi}{O_J}{\Psi}=0$. Indeed, it is easy to check that the distance between a closest pair of vertices respectively appearing in two separated loops among the four with broken constraints in $\ket{\phi_m}$ is $(\mathcal{D}-1)/2$, greater than $\mathcal{D}_J$ so that $J$ can only cover vertices in one such loop. Consequently, $\mel{\psi_{m'}}{O_J}{\phi_m}=0$ since $O_J$ keeps $O_J\ket{\phi_m}$ with at least one constraint broken and different from that in $\ket{\psi_{m'}}$. Hence we have proved $\mel{\Psi}{O_J}{\Phi}=0$, and $\mel{\Phi}{O_J}{\Psi}=0$ as well using the same argument.

To prove $\expval{O_J}{\Psi}=\expval{O_J}{\Phi}$, we define unitary operators 
\begin{align*}
\begin{split} 
S^1_{i}=\dyad{0}{1}+\dyad{1}{0}+\dyad{2}{3}+\dyad{3}{2},\\
S^2_{i}=\dyad{0}{2}+\dyad{2}{0}+\dyad{1}{3}+\dyad{3}{1},\\
S^3_{i}=\dyad{0}{3}+\dyad{3}{0}+\dyad{1}{2}+\dyad{2}{1},
\end{split}
\end{align*}
acting on a single qudit located at the vertex $i$. As illustrated in Fig.~4a and 4d in the main text, the effect of these operators is to exchange the red and pink on two sides of the triangle or create two red (pink) sides from two pink (red) sides. Then, since vertex $i$ simultaneously appears in three loops (including the laterals), applying $S^1_{i},S^2_{i}$ or $S^3_{i}$ to $\ket{\psi_m}$ simply changes the constraints to the opposite on two out of the three loops according to the superscript $1,2,3$. With these unitary operators, we can define a unitary operator $V_1=S^1_{i_3}S^1_{i_2}S^1_{i_1}$ where each of the three vertices $i_1,i_2,i_3$ appears in two out of the four largest loops. As shown in Fig.~4a and 4b in the main text, $V_1$ simply changes the constraints on the four loops and maps $\ket{\psi_m}$ to $\ket{\phi_{m'}}$ and hence maps $\ket{\Psi}$ to $\ket{\Phi}$.

Now, there are two cases to be considered regarding the support $J$ of $O_J$: (1) $J$ does not cover any of $i_1,i_2,i_3$; (2) $J$ covers only one of the three vertices. Indeed, as we have shown, $J$ cannot cover more than one of the three vertices since they are farther apart beyond the size $\mathcal{D}_J$. In case (1), we have $[O_J,V_1]=[O_J,V^+_1]=0$. It follows that $\expval{O_J}{\Phi}=\expval{O_J}{V_1\Psi}=\expval{V^+_1O_JV_1}{\Psi}=\expval{V^+_1V_1O_J}{\Psi}=\expval{O_J}{\Psi}$ as desired. In case (2), without loss of generality, we suppose the covered vertex is $i_1$. Then we can define another two unitary operators 
\begin{align*}
\begin{split} 
V_2&=S^1_{i_3}S^1_{i_2}S^2_{k_{(\mathcal{D}+1)/4}}\cdots S^2_{k_2}S^3_{k_1},\\
V_3&=S^1_{i_3}S^1_{i_2}S^3_{k'_{(\mathcal{D}+1)/4}}\cdots S^3_{k'_2}S^2_{k'_1},
\end{split}
\end{align*}
where the two subsets of vertices $K=\{k_1,k_2,\ldots,k_{(\mathcal{D}+1)/4}\}$ and $K'=\{k'_1,k'_2,\ldots,k'_{(\mathcal{D}+1)/4}\}$, as specified in Fig.~4d in the main text form the two paths connecting two adjacent largest loops. Note that any $k\in K$ and $k'\in K'$ have distance $(\mathcal{D}+1)/4,(\mathcal{D}+1)/4-1$ to $i_1$ respectively. Then, since $i_1$ is covered in $J$, $J$ cannot have overlap with both $K$ and $K'$, otherwise the distance between the $k$ and $k'$ (in the overlap) within $J$ exceeds the size $\mathcal{D}_J$, which implies contradiction. Indeed, due to the inequality $\mathcal{D}_J\le\frac{3}{8}\mathcal{D}-\frac{5}{8}$, $J$ cannot cover the whole of or encircle any of the four loops, since any connected region encircling such a loop has size at least equal to $\frac{3}{8}\mathcal{D}+\frac{3}{8}>\mathcal{D}_J$. Hence, if $k$ and $k'$ are both in $J$, the shortest path connecting them (which defines their distance with $J$) must pass through $i_1$, giving rise to the distance $\frac{1}{2}\mathcal{D}-\frac{1}{2}>\mathcal{D}_J$. Consequently, we have either $[O_J,V_2]=[O_J,V^+_2]=0$ or $[O_J,V_3]=[O_J,V^+_3]=0$, which leads to $\expval{O_J}{\Psi}=\expval{O_J}{\Phi}$ by the same arguments as in case (1).

Above arguments have proved the error-detecting property, and that any $U$ characterized by given $P$ and $L$ cannot completely disentangle $\ket{\Psi}$ once the system size is large enough. Together with the zero correlation length proved in the previous subsection, we have proved the LRE in $\ket{\Psi}$. We have also proved that for given $P$, the depth $L$ for any local quantum circuit characterized by $P$ and $L$ to completely disentangle $\ket{\Psi}$ has a lower bound $L>\frac{3}{16P}\mathcal{D}+\frac{1}{2}-\frac{5}{16P}$ which is linear to the lattice linear size $\mathcal{D}$.

\begin{figure}[ht]
\centering
    \includegraphics[width=7.5cm]{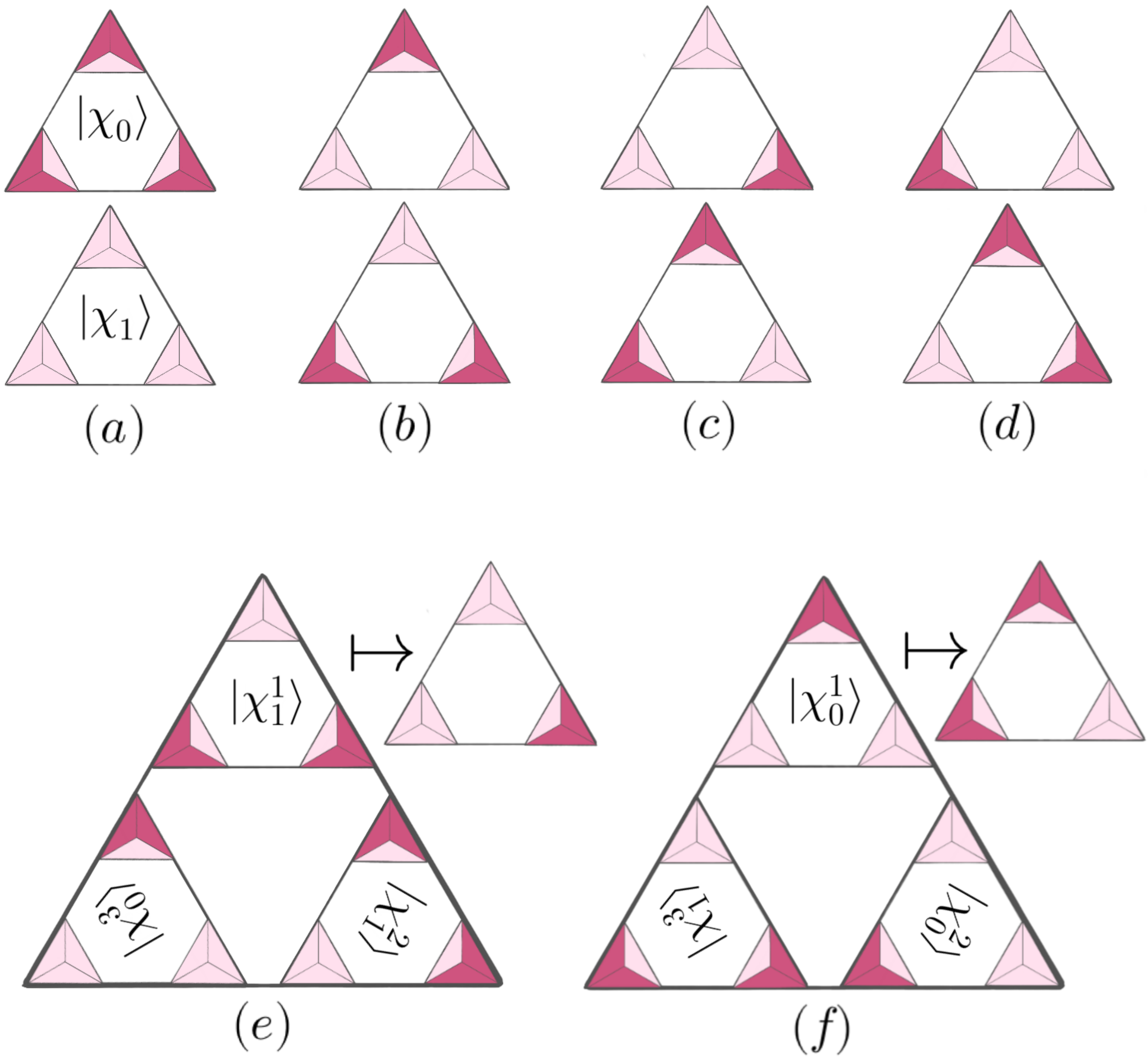}   
\phantomsubfloat{\label{5a}}\phantomsubfloat{\label{5b}}
\phantomsubfloat{\label{5c}}\phantomsubfloat{\label{5d}}
\phantomsubfloat{\label{5e}}\phantomsubfloat{\label{5f}}
\vspace{0\baselineskip}
\caption{(a), (b), (c) and (d) are the four possible cases for $\ket{\chi_0}$ (upper) and $\ket{\chi_1}$ (lower). (e) and (f) are the two combination $\ket{\chi^1_1}\otimes\ket{\chi^2_1}\otimes\ket{\chi^3_0}$ and $\ket{\chi^1_0}\otimes\ket{\chi^2_0}\otimes\ket{\chi^3_1}$ that $\ket{\chi}$ can be mapped to, in which each pair of neighboring qudits linking the big blocks is either in $\ket{0}\otimes\ket{0}$ or in $\ket{1}\otimes\ket{1}$. The two states are eventually mapped to the desired $\ket{\chi_0}$ and $\ket{\chi_1}$.}
\end{figure}

\section{The proof for the ergodicity of the $T_{ii'}$ operators}
The consideration mentioned in the main text is that by definition, all $T_{ii'}$ operators commute with each other, and they are all unitary and satisfy $T_{ii'}\ket{\Psi}=\ket{\Psi}$. The effect of each $T_{ii'}$ simply maps one $\ket{\psi_m}$ to another $\ket{\psi_{m'}}$, both satisfying the constraints on all loops. Then, since $\ket{\Psi}$ is an equal-weight summation, ergodicity of $T_{ii'}$ operators will lead to the formula $\ket{\Psi}\propto\prod_{ii'}\frac{\mathds{1}+T_{ii'}}{\sqrt{2}}\ket{000\cdots}$. Note that $\ket{000\cdots}=\ket{\psi_{m_0}}$ also satisfies all constraints.

To prove the ergodicity, it suffices to prove that any $\ket{\psi_m}$ can be mapped to $\ket{\psi_{m_0}}=\ket{000\cdots}$ by applying the $T_{ii'}$ operators finite times on different pairs of neighboring vertices $(i,i')$.

In this section we prove a more general condition which directly implies the desired condition. We consider a qudit-product-state $\ket{\chi}$ of the form $\ket{\alpha\alpha'\alpha''\cdots}$. For convenience in our proof, we reorganize the basis states in the qudit-product form as $\ket{\alpha_1\alpha_2\alpha_3}\otimes\ket{\alpha\alpha'\alpha''\cdots}$ where $\alpha_1,\alpha_2,\alpha_3$ stand for the single-qudit states on the three corner vertices while $\alpha\alpha'\alpha''\cdots$ stand for the rest. We still use the colored triangle to represent each $\ket{\alpha}$, and assume that $\ket{\chi}$ satisfies the same constraints (even number of red sides) on all loops but not on the laterals of the lattice.

We want to prove the following condition: By applying the $T_{ii'}$ operators enough times on certain pairs of neighboring vertices $(i,i')$, $\ket{\chi}$ can be mapped to exactly two qudit-product-states $\ket{\chi_0}$ and $\ket{\chi_1}$ of the form 
$\ket{\alpha_1\alpha_2\alpha_3}\otimes\ket{000\cdots}$, i.e. with all qudits in $\ket{0}$ except for the three corners; Furthermore, $\ket{\chi_0}$ and $\ket{\chi_1}$ must be one of the four cases illustrated in Fig.~\ref{5a}, \ref{5b},  \ref{5c} and \ref{5d} respectively where we omit the states $\ket{000\cdots}$ and only represent $\ket{\alpha_1\alpha_2\alpha_3}$ pictorially for simplicity.

We prove by induction on the system size, i.e. the finite generation of the Sierpi\'nski lattice. It is trivial to prove that the condition holds for the smallest generation, i.e., when states are defined on a block. Now we assume that the condition is true for the $n$-th generation. We just need to prove that the condition holds for the $n+1$-th generation.

Consider $\ket{\chi}$ on the lattice of the $n+1$-th generation. Since $\ket{\chi}$ is a qudit-product-state, it can always be written as $\ket{\chi}=\ket{\chi^1}\otimes\ket{\chi^2}\otimes\ket{\chi^3}$ where $\ket{\chi^1}$, $\ket{\chi^2}$ and $\ket{\chi^3}$ are the states on the three big triangular blocks which together form the whole lattice. Obviously, $\ket{\chi^1}$, $\ket{\chi^2}$ and $\ket{\chi^3}$ are on the lattice of the $n$-th generation, satisfying the constraints on all loops therein, and hence can be mapped to $\ket{\chi^1_0},\ket{\chi^1_1}$, $\ket{\chi^2_0},\ket{\chi^2_1}$ and $\ket{\chi^3_0},\ket{\chi^3_1}$ respectively. In other words, by applying the $T_{ii'}$ operators enough times on certain pairs of neighboring vertices $(i,i')$, $\ket{\chi}$ can be mapped to the eight combination of the six states, i.e. $\ket{\chi^1_0}\otimes\ket{\chi^2_0}\otimes\ket{\chi^3_0}, \ket{\chi^1_1}\otimes\ket{\chi^2_1}\otimes\ket{\chi^3_1},\ldots$. Note that in any of such combination, there are even number of $\ket{1}$ on the six qudits connecting the three big blocks, since $\ket{\chi}$ satisfies the constraints (even number of red sides) on the largest loop. It can be easily shown that there are exactly two of such combinations, for example $\ket{\chi^1_1}\otimes\ket{\chi^2_1}\otimes\ket{\chi^3_0}, \ket{\chi^1_0}\otimes\ket{\chi^2_0}\otimes\ket{\chi^3_1},\ldots$ as illustrated in Fig.~\ref{5e} and \ref{5f} respectively, in which each pair of neighboring qudits linking the big blocks is either in $\ket{0}\otimes\ket{0}$ or in $\ket{1}\otimes\ket{1}$. Then, as shown in Fig.~\ref{5e} and \ref{5f}, applying the $T_{ii'}$ to those pairs in $\ket{1}\otimes\ket{1}$, $\ket{\chi}$ is eventually mapped to two states of the form $\ket{\alpha_1\alpha_2\alpha_3}\otimes\ket{000\cdots}$. Furthermore, according to the four cases illustration in Fig.~\ref{5a}, \ref{5b},  \ref{5c} and \ref{5d}, the eventual two states are exactly within one of the four cases, and hence are the desired $\ket{\chi_0}$ and $\ket{\chi_1}$. Indeed, according to the constraints of $\ket{\chi}$ on the laterals, $\ket{\chi_0}$ and $\ket{\chi_1}$ and the only possible cases of the form $\ket{\alpha_1\alpha_2\alpha_3}\otimes\ket{000\cdots}$. Therefore, the desired condition is proved.

Now, if we apply the condition to the case of consideration, it is obvious that any $\ket{\psi_m}$ can be mapped to $\ket{\psi_{m_0}}=\ket{000\cdots}$ by applying the $T_{ii'}$ operators finite times on different pairs of neighboring vertices $(i,i')$.

\end{document}